\documentclass[11pt,twoside]{article}

\usepackage{asp2004}
\usepackage{epsf}
\usepackage{psfig}
\usepackage{lscape}

\def\secspt{$\buildrel{\prime\prime}\over .$}

\begin{document}
\title{ Star-Cluster Astrometry with Ground-Based Wide Field Imagers} 
\author{Luigi R.\ Bedin}   
\affil{ESO - Garching, Germany, EU; \\lbedin@eso.org}    
\author{Jay Anderson}   
\affil{Dept. of Physics and Astronomy, Rice Univ., Houston, TX, USA; jay@eeyore.rice.edu}    
\author{Giampaolo Piotto, Yazan Momany, Ramakant S.\ Yadav}   
\affil{Dip. di Astronomia, Univ. di Padova, Padova, Italy, EU; piotto-momany-yadav@pd.astro.it}    

\begin{abstract} 
We  show the astrometric  potential of  the Wide  Field Imager  at the
focus  of the  MPI-ESO  2.2m  Telescope.  Currently,  we  are able  to
measure  the position  of  a  well-exposed star  with  a precision  of
$\sim$4  mas/frame  in  each   coordinate  (under  0.8  arcsec  seeing
conditions).  We present some preliminary results here.
\end{abstract}


%
\section{Introduction}
In the last few years, several  Wide Field Imagers (WFIs) at the focus
of large ground-based telescopes  have become operative (MPI-ESO 2.2m,
AAT  4m, CFH  4m), and  their number  is continuously  increasing (LBT
2$\times$8m, VST 2.5m,  UKIRT 3.8m, VISTA 4m, etc.),  as well as their
field of view.  These WFIs have  allowed us to map completely a number
of  open and  globular clusters  in our  Galaxy, and  to  get accurate
photometry for large numbers of stars, with the additional possibility
of studying fast-evolving phases of stellar evolution.
By definition, the  WFIs allow large radial coverage  in a cluster, so
that  we can  study  the  radial distribution  of  stars in  different
sequences  of  the  color--magnitude  diagram (CMD)  and  of  peculiar
objects, which allows us to  investigate the effect of the environment
on the  evolution of the cluster  stars.  The wide  field coverage has
made the study of tidal tails  in open and globular clusters much more
practicable.

Among the  most interesting opportunities  offered by the  WFIs (still
largely  unexplored) are in  their astrometric  performance.  Accurate
astrometry  over wide  fields is  important for  a number  of reasons.
First of all, an accuracy of 0.2 arcsec or better is required to point
the   fibers   of    multi-fiber   spectroscopic   facilities   (e.g.,
FLAMES$+$GIRAFFE/VLT at ESO).
However,  the most  promising  applications lie  in the  proper-motion
measurements  of a  large  number of  stars.   WFIs allow  astrometric
measurements with an accuracy of far better than $0.2$ arcsec.  With a
baseline of a few years, images collected with modern WFIs can provide
proper  motions more accurate  than those  obtainable with  old plates
with a baseline of several  decades.  (Note, though, that these plates
will still  remain valuable for long-term  non-linear astrometry, such
as the determination of the  orbit of long-period visual binaries, and
of course for long-term variation in the light curves).
%
\begin{figure}
\plottwo{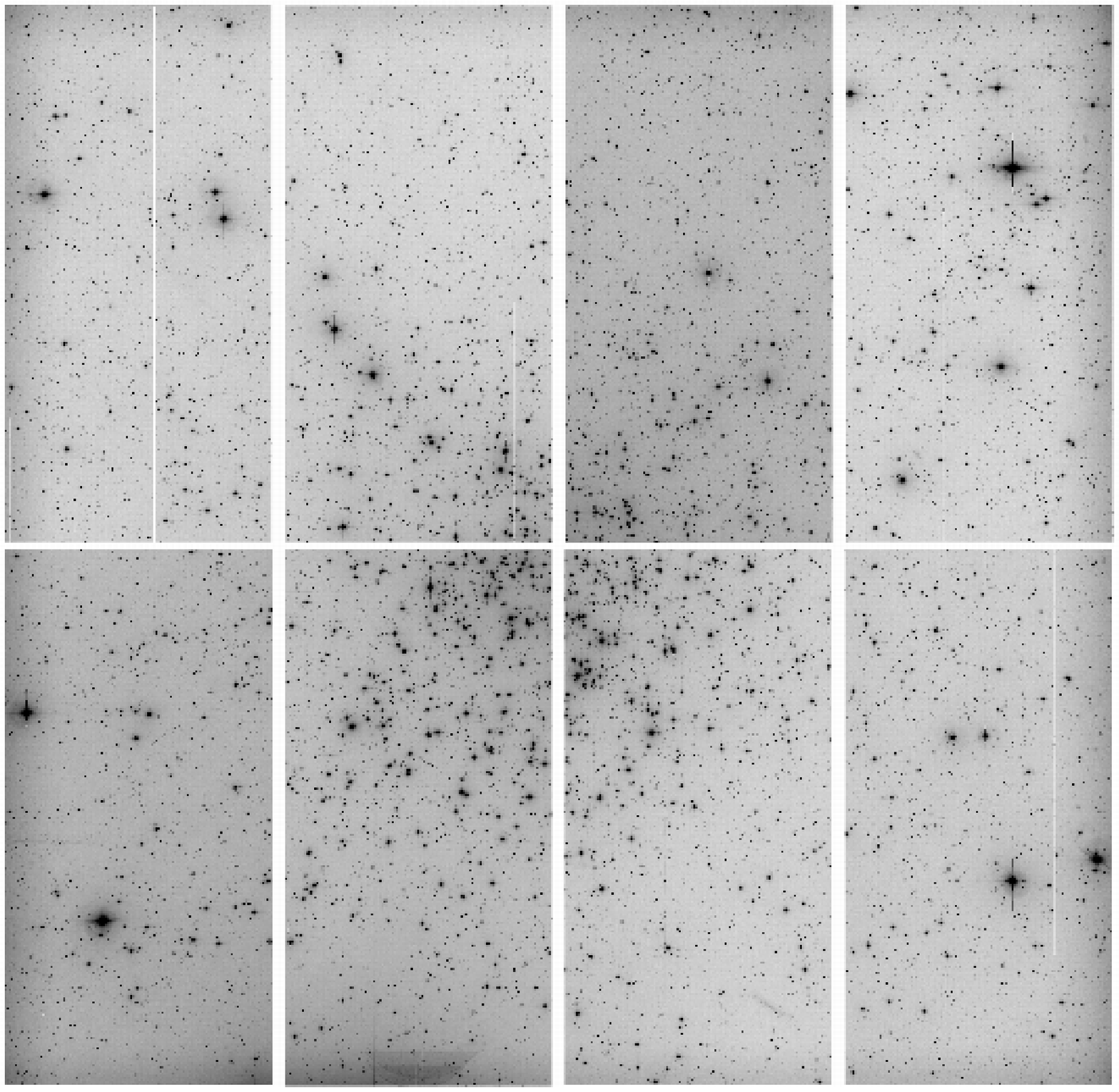}{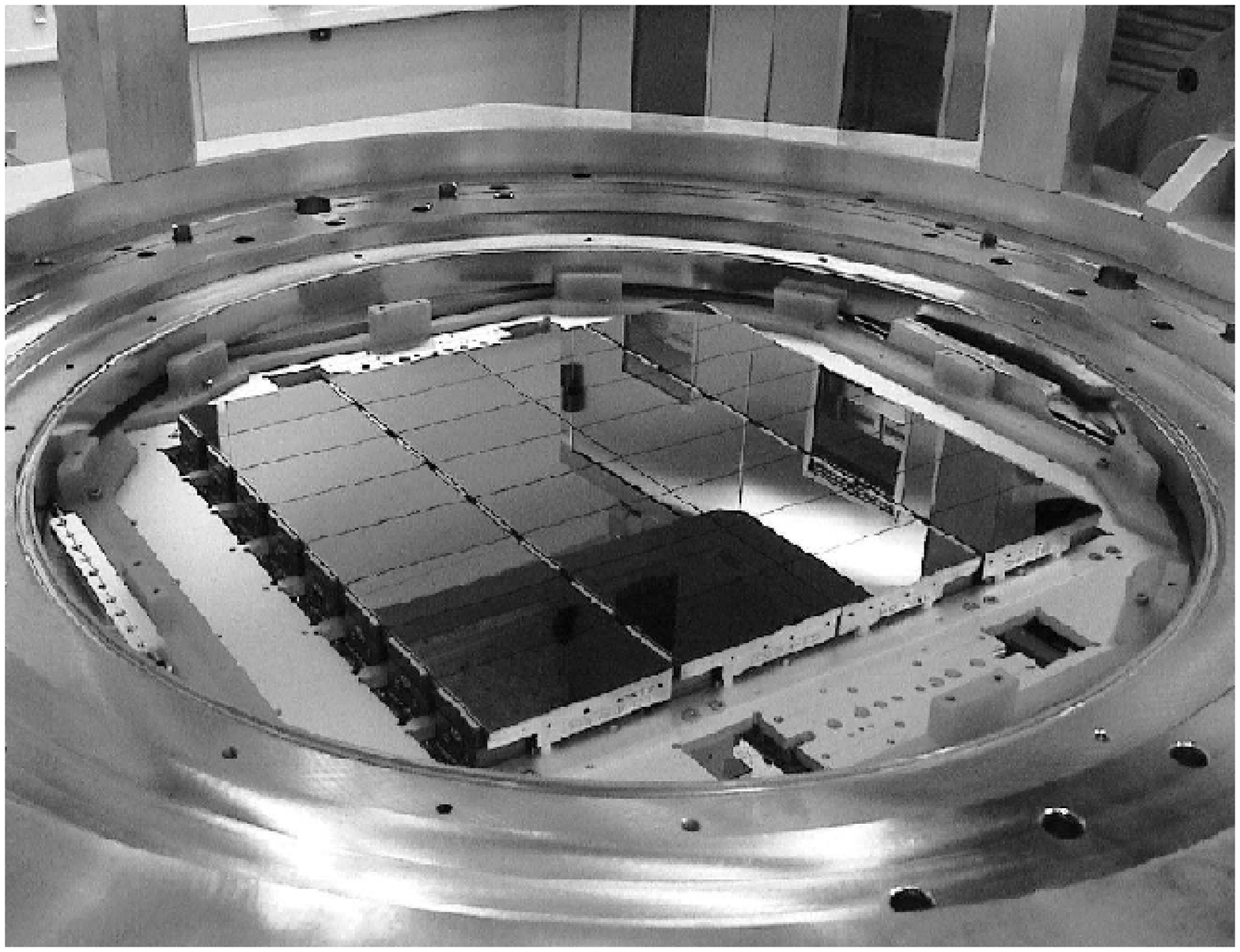}
\caption{On  the  {\em  left},  an  image from  the  WFI/MPI-ESO  2.2m
(WFI@2.2m), centered  on the Galactic  open cluster NGC 2477.   On the
{\em  right}, OMEGACAM,  a  WFI four  times  as large  (both in  field
coverage, and  in number of pixels).  OMEGACAM will be  mounted at the
beginning of  2006 at the focus  of the VST-ESO  2.5m telescope.  This
telescope has  been specifically designed for wide  field imaging, and
we we  expect that  OMEGACAM will provide  better astrometry  than the
WFI@2.2m. }
\end{figure} 
%
\section{Astrometry: The importance of a careful  PSF modeling}
In the last year, we have  started to apply to wide field ground-based
images  what  we have  learned  from  Hubble  Space Telescope  ($HST$)
(Anderson \& King 2000, 2003, Anderson 2004).
For accurate astrometry  the ability to reproduce the  the core of the
Point Spread  Function (PSF)  is of crucial  importance, much  more so
than  for accurate  PSF-fitting  photometry.  In  fact,  the PSF  core
(where the  derivatives of the  stellar profile are  highest) contains
almost  all the  astrometric information.  The  PSF core  needs to  be
carefully  modeled, as  does its  dependence on  the  spatial position
within  the  detector. Moreover,  the  core of  the  PSF  needs to  be
represented with adequate sampling.
In  Figure 2  we  show  the precision  achieved  in measuring  stellar
positions in images collected  at the WFI@2.2m, under 1\secspt2 seeing
conditions.   We verified  that  a seeing  of  0\secspt8 improves  the
precision by $\sim$30\%.

At the moment our main  limitation comes from the geometric distortion
of the focal plane, and we  are working on an algorithm to correct for
it.  Before we  can correct for it, we must  of course understand what
the  nature of  the  distortion is,  in  terms of  (1)  what order  of
polynomial best characterizes it,  (2) whether there is any fine-scale
component  added by  the filters  or  other optical  elements and  (3)
whether  it   changes  over  time.    Even  if  we   cannot  perfectly
characterize  the distortion,  it is  still possible  to  minimize its
effect on astrometry either by taking  a set of exposures at a variety
of pointing  offsets or by doing  transformations in a  more local way
(Bedin et al.\ 2003).
%
\begin{figure}
\plotone{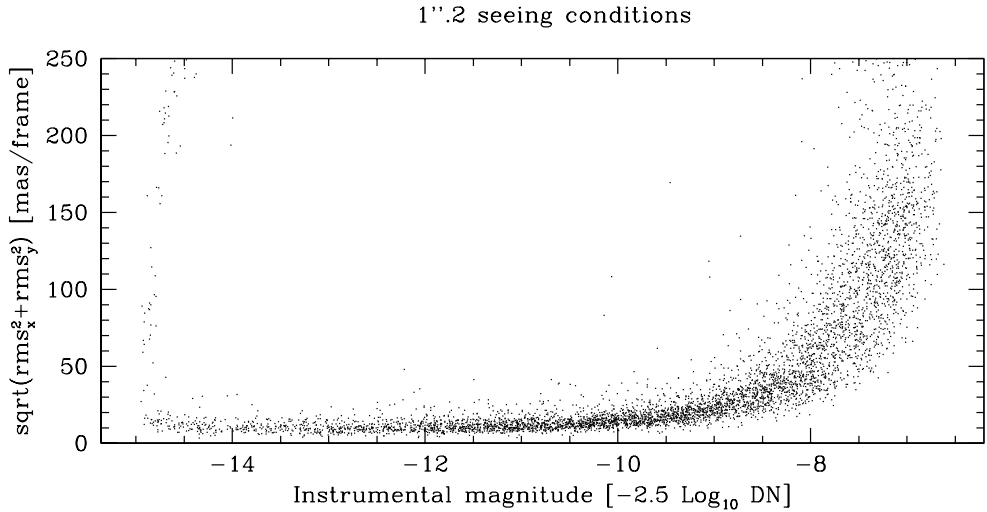}
\caption{Errors   in  measuring   star  positions   vs.\  instrumental
magnitude,  in a  sample  of WFI@2.2m  images  collected under  seeing
conditions  of 1\secspt2.   The instrumental  magnitude is  defined as
$-2.5 \log_{10}$ of the Digital Numbers (DN). }
\end{figure} 
%
\section{Example: M4 field decontamination in 2.2 years}
Figure  3   shows  an  example  of  the   proper-motion  potential  of
ground-based  wide field  imagers; we  present preliminary  results on
field-star  removal  in  part  of the  low-Galactic-latitude  globular
cluster  M4.  Observations  collected at  the WFI@2.2m  in  two epochs
separated  by a  time  baseline of  just  2.2 years  already allow  an
excellent separation.

The first-epoch  data consist of 3$\times$75s  + 2$\times$55s $B$-band
images taken on December 6, 1999, and the second-epoch data consist of
3$\times$180s  images in  the  same band,  and 3$\times$120s  $V$-band
images, taken on February 19, 2002.
In order to avoid first-order color  terms, we use only the $B$ images
to derive the proper motions, and  we use only the cluster MS stars as
reference stars in  the transformation, so that stars  moving with the
cluster will have zero displacement.
The top left  panel of Figure 3 shows the vector  point diagram of the
displacements, in units of WFI pixels (238 mas/pixel).
From high-accuracy  astrometric measurements  on $HST$ data,  Bedin et
al.\  (2003) have  shown that  the average  proper-motion displacement
between cluster stars and field objects is $\sim$17 mas/yr.
Since   our  astrometric  errors   increase  rapidly   toward  fainter
magnitudes, we consider as cluster members all the stars with a proper
motion which  differs by less than  10 mas/yr from  the average proper
motion of the cluster MS stars.

This example must  be considered as an application  that is still very
preliminary, but  it shows the importance  of high-accuracy astrometry
on wide field ground-based images.

\begin{figure}
\plotfiddle{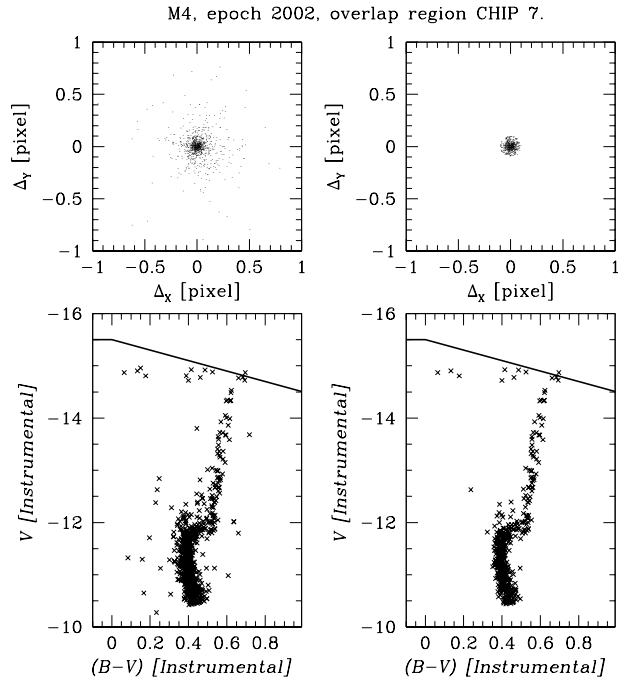}{8.9truecm}{0}{45}{45}{-150}{-60}
\caption{Proper-motion measurements  in a field of M4,  using two sets
  of WFI@2.2m images separated by only 2.2 yr.  }
\end{figure} 

\acknowledgements 
We thank Ivan R.\ King for a careful reading of the manuscript. 


\end{document}